\documentclass[aps,epsfig,floats,prl,twocolumn,showpacs]{revtex4-1}
\usepackage{graphicx}

\begin{document}
\title{Synthetic Mechanochemical Molecular Swimmer}
\author{Ramin Golestanian}
\email{r.golestanian@sheffield.ac.uk}

\affiliation{Department of Physics and Astronomy, University of
Sheffield, Sheffield S3 7RH, UK}

\affiliation{Laboratoire de Physico-Chimie Th\'eorique, UMR CNRS
Gulliver 7083, ESPCI, 75231 Paris Cedex 05, France}

\date{\today}

\begin{abstract}
A minimal design for a molecular swimmer is proposed that is a based
on a mechanochemical propulsion mechanism. Conformational changes
are induced by electrostatic actuation when specific parts of the
molecule temporarily acquire net charges through catalyzed chemical
reactions involving ionic components. The mechanochemical cycle is
designed such that the resulting conformational changes would be
sufficient for achieving low Reynolds number propulsion. The system
is analyzed within the recently developed framework of stochastic
swimmers to take account of the noisy environment at the molecular
scale. The swimming velocity of the device is found to depend on the
concentration of the fuel molecule according to the Michaelis-Menten
rule in enzymatic reactions.
\end{abstract}


\maketitle

One of the aims of nanotechnology is to be able to make synthetic
molecular devices that could propel themselves through fluidic
environments and perform targeted tasks such as delivery of
therapeutic agents or carrying out mechanical work
\cite{kay}. At such small scales, one cannot apply
standard deterministic strategies used in engineering at the
macroscopic scales, as the dynamics of any device will be
overwhelmed by fluctuations of thermal or other origins. The right
strategy is to find a way to impose a bias on these fluctuations
such that they will {\em average out} to our desired behavior, as
can be learned from the example of biological molecular motors
\cite{armand}.

In addition to enduring the noisy environment, molecular swimmers
would also need to overcome the fundamental difficulty posed by the
governing hydrodynamic rules at low Reynolds number conditions: they
should utilize a non-reciprocal sequence of deformations to break
the time reversal symmetry in their periodic motions and achieve
swimming \cite{purcell}. This is a nontrivial
task if one is to use a minimal set of degrees of freedom
\cite{becker,Ali,josi,drey,feld,peko,yeomans1,Karsten}, which will
most likely be what can be afforded in a human-engineered device.

There has been a significant recent development along this line with
the advent of a number of experimentally made micron-scale
prototypes of such low Reynolds number swimmers. Using magnetic
actuation of micron-sized linked magnetic beads
\cite{dreyfus05,Pietro} or manipulation of particles using optical
tweezers \cite{cicuta}, it was demonstrated that low Reynolds number
propulsion can be achieved at the micro-scale via non-reciprocal motion.
In these experiments, the actuation mechanisms were externally enforced
(using oscillating magnetic fields or optical traps), which means that
strictly speaking the devices were not self-propelled swimmers. Moreover,
it is not clear if such actuation mechanisms could be scaled down to be
applicable to molecular devices. This means that we are in need of
new strategies to be able to program non-reciprocal molecular
deformations.

While other external triggers---such as laser pulses that could
induce conformational changes---could in principle be pursued
\cite{kay}, it would be ideal to be able to couple the
conformational changes to a local source of free energy via a
chemical reaction, like motor proteins \cite{howard}. This will
allow the mechanism to potentially accommodate arbitrary degrees of
complexity, and avoid interference between different parts of the
system. To my knowledge, this scenario has not been considered so far in the
literature for microscopic swimmers, although there have been demonstrations of
mechanical actuation of elastic gels by using oscillatory chemical reactions
(at macroscopic scales) \cite{gel}.
A strategy based on catalysis of a chemical reaction has
already been used for making self-propelled colloidal particles,
which utilize non-equilibrium interfacial (phoretic) interactions
towards their net motion \cite{phoretic,jon}. While this mechanism 
of motility has shown to be extremely powerful, exploring other possible
strategies would help create a diverse range of approaches that could be 
used in the future to optimize devices for specific purposes and overcome
strategy-specific hurdles for a particular end use. For example, a swimmer
that utilizes interfacial interactions as its main motility mechanism is
potentially vulnerable to intervention by other interfacial forces when it
is close to boundaries or other devices. Moreover, one wonders why nature has 
not chosen to take advantage of the ``phoretic'' strategy in any living 
system; one may speculate that it could be because it cannot be conveniently 
miniaturized and robustly function in maximally crowded intracellular environments.

Here I propose an alternative minimal design for a molecular swimmer, which
would be able to catalyze chemical reactions and use the free
energy gain and products to induce the conformational changes that
would be sufficient for achieving low Reynolds number propulsion in
a noisy environment \cite{note0}. The present study could also shed some 
light on how the motor proteins could combine enzymatic action with conformational 
changes, as it presents a full physical analysis of a simple model that explains 
how such a coupling could work in practice.

\begin{figure}
\begin{center}
\includegraphics[width=0.7\columnwidth]{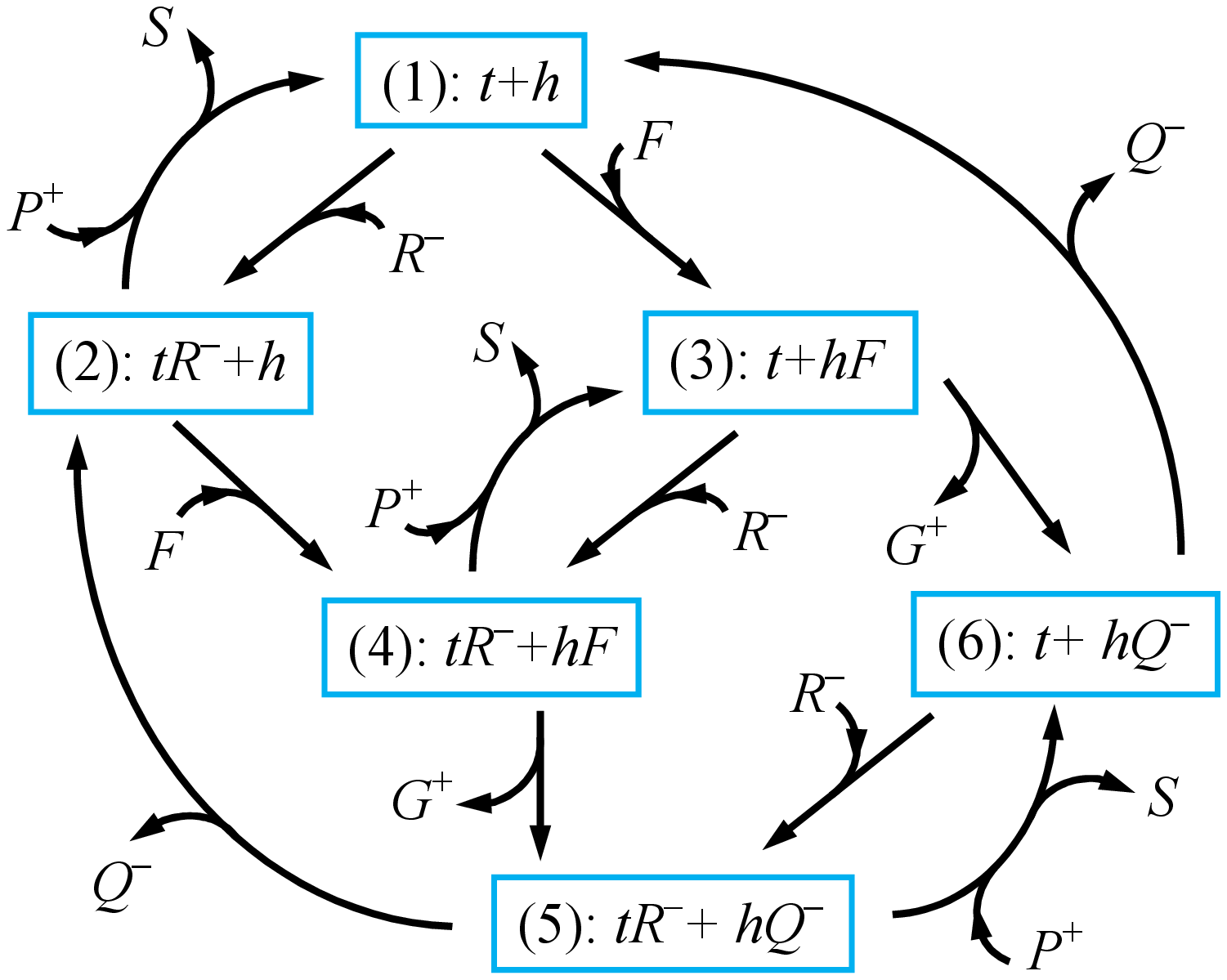}
\end{center}
\caption{The 6-state chemical cycle of the three-sphere
molecular device that comprises enzyme $h$ at the head
and enzyme $t$ at the tail (see Fig. \ref{fig:states}).
$h$ catalyzes $F \rightarrow Q^{-}+G^{+}$ and
$t$ catalyzes $R^{-}+P^{+} \rightarrow S$.}
\label{fig:cycle}
\end{figure}

The proposed design for the molecular swimmer
%that could couple conformational changes with chemical reactions
is based on the simple three-sphere low Reynolds number swimmer model \cite{Ali,RamArm08}.
To incorporate catalytic activity, we take two enzymes, $h$ and $t$,
and place them at the {\em head} and the {\em tail} of the three-sphere device,
respectively. The enzymes catalyze the following chemical reactions
\begin{eqnarray}
&& F+h \rightarrow hF \rightarrow hQ^{-}+G^{+} \rightarrow h+Q^{-}+G^{+},\label{eq:h-reac} \\
&& R^{-}+P^{+}+t \rightarrow tR^{-}+P^{+} \rightarrow t+S,\label{eq:t-reac}
\end{eqnarray}
which would otherwise occur very slowly in the bulk. Since the head can be
found in three distinct states of $h$, $hF$, and $hQ^{-}$, and the tail
has two distinct states of $t$ and $tR^{-}$, the combined system can have 6 distinct
chemical states. The reactions (\ref{eq:h-reac}) and (\ref{eq:t-reac}) then multiply
within these states and produce the chemical cycle shown in Fig. \ref{fig:cycle}.

\begin{figure}
\begin{center}
\includegraphics[width=1.00\columnwidth]{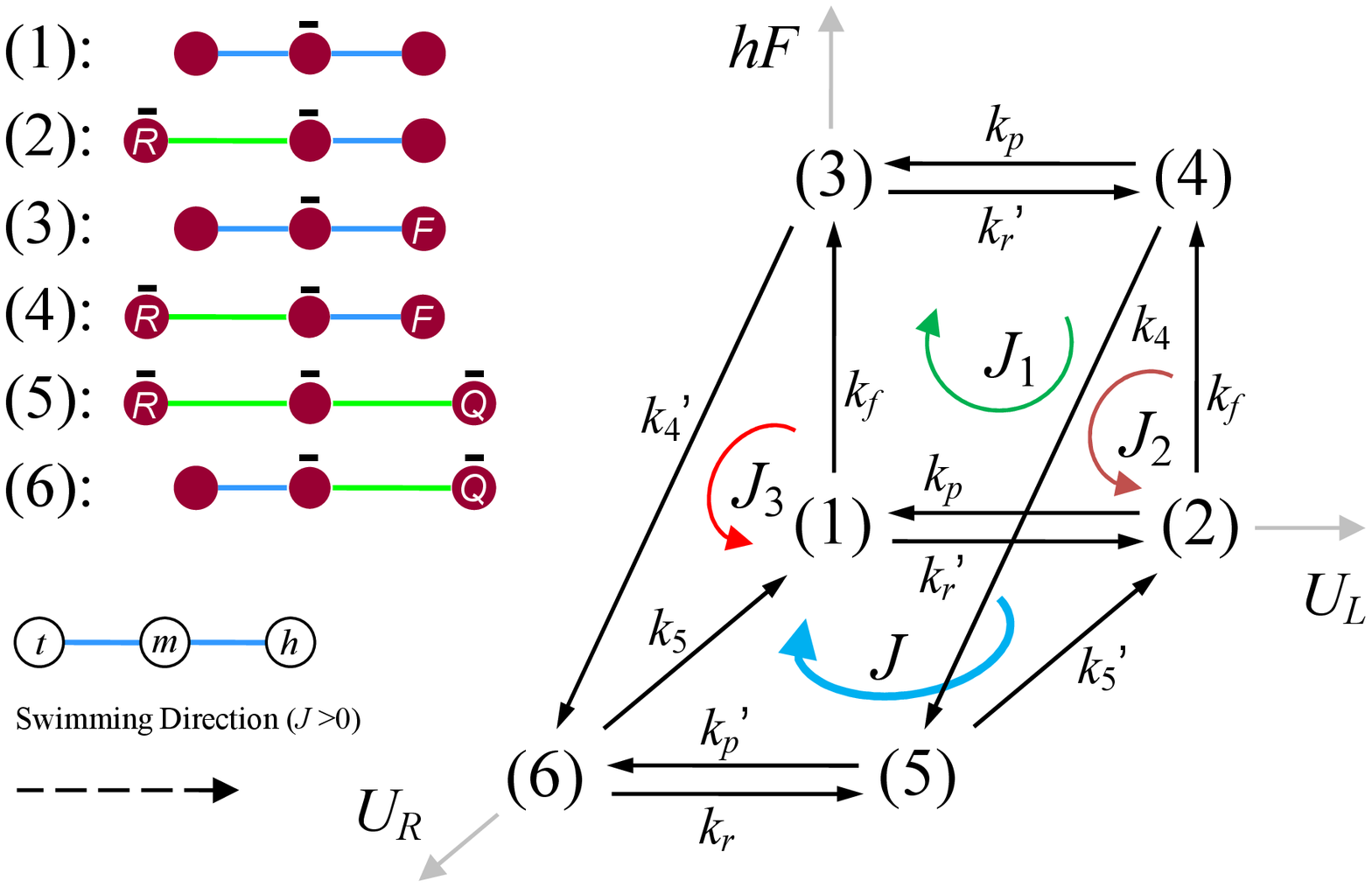}
\end{center}
\caption{The 6 distinct mechanochemical states of the three-sphere
swimmer in the configuration space. 
%The middle sphere is negatively charged and binding of negatively 
%charged ions at either head or tail of the swimmer leads to an 
%elongation of the corresponding arm due to electrostatic repulsion. 
The conformational states of the swimmer are determined by the projection 
onto the space of the deformation of the left ($U_L$) and right ($U_R$) arms, 
while binding of $h$ with $F$ produces distinct chemical states that have 
identical conformations. All possible transitions described in the chemical 
cycle in Fig. \ref{fig:cycle} are shown by arrows with attributed rate constants.
}
\label{fig:states}
\end{figure}

The second ingredient in the design is to take advantage of the presence of
ionic components in the reactions and use electrostatic interactions to induce
conformational changes in the swimmer. If we make the {\em middle} sphere
permanently negatively charged (see Fig. \ref{fig:states}), the temporarily
charged states of $hQ^{-}$ and $tR^{-}$ will introduce electrostatic repulsion
that can elongate the right and left arms, respectively. This allows us to
map the 6 chemical states shown in Fig. \ref{fig:cycle} onto the 6 distinct
mechanochemical states shown in Fig. \ref{fig:states}. These mechanochemical
states can be represented in a 3D configuration space whose axes correspond
to deformation of the right arm $U_R$, deformation of the left arm $U_L$,
and complexation of $h$ with $F$. The reaction routes in the chemical cycle
in Fig. \ref{fig:cycle} can now be used to identify the corresponding transitions
between the mechanochemical states in the configuration space, as shown in
Fig. \ref{fig:states}. The relevant transition rates are defined on the
configuration space diagram in Fig. \ref{fig:states}. I have chosen a
notation where the rates of similar transitions are differentiated by a prime,
with the general rule that $k_{\alpha}'>k_{\alpha}$. These choices are made
depending on whether the presence of additional electrostatic charges hampers or
facilitates a certain reaction \cite{note1}.

\begin{figure*}
\begin{center}
\includegraphics[width=2.00\columnwidth]{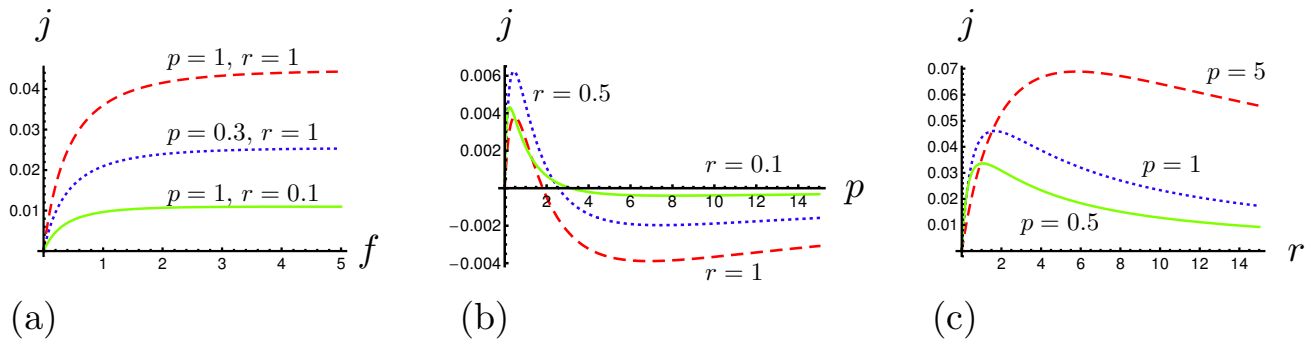}
\end{center}
\caption{Dimensionless current $j=J/k_5$ as a function of $f=k_{f} [F]/k_5$,
$r=k_{r} [R^{-}]/k_5$, and $p=k_{p} [P^{+}]/k_5$, corresponding to $\alpha=1.5$,
$\beta=1.5$, $\delta=1.1$, and $b=1$. (a) Michaelis-Menten behavior, for
$\gamma=1.1$. (b) Sign change, for $f=1$ and $\gamma=2.7$. (c) Nonmonotonic
concentration dependence, for $\gamma=1.1$ and $f \to \infty$.}
\label{fig:plots}
\end{figure*}

We do not make assumptions about any specific ordering of the
various events involved in the cycle (see Figs. \ref{fig:cycle} and \ref{fig:states}).
For example, state $(1)$ can proceed to state $(2)$
upon binding of $R^{-}$ to $t$, and can also proceed to $(3)$ if $F$ binds
to $h$. Any of these two can happen at a time in a stochastic
manner, and only the average rates of these transitions are controlled.
Moreover, not all of the transitions contribute constructively towards net swimming.
For example, state $(2)$ can proceed to state $(4)$ upon binding
of $F$ to $h$, which will take the system further down towards completing
a non-reciprocal mechanical cycle, but could also go back to state $(1)$
if $P^{+}$ binds to $tR^{-}$, which undoes the progress made
by the previous transition towards net swimming.

The dynamics of the device can be studied using a stochastic
description of the conformational transitions and the relative
population of the different states at stationary state
\cite{RamArm08}. In stationary situation, each of the six mechanochemical
states shown in Fig. \ref{fig:states} is populated with a probability
$p_i$ ($i=1,\cdots,6$), subject to the normalization condition
$p_1+p_2+p_3+p_4+p_5+p_6=1$. The loops in the configuration space
contribute probability currents denoted as $J$, $J_1$, $J_2$, and $J_3$
in Fig. \ref{fig:states}. Conservation of probability current leads to
the following nine equations between the probabilities and the currents:
$J+J_3=k_5 p_6$, $J_3+J_1=k_f [F] p_1$, $J_3=k_{4}' p_3$, $J_2-J=k_5' p_5$,
$J_2-J_1=k_f [F] p_2$, $J_2=k_{4} p_4$, $J_1=k_{r}' [R^{-}] p_3-k_{p} [P^{+}] p_4$,
$J=k_{p}' [P^{+}] p_5-k_{r} [R^{-}] p_6$, $J-J_1=k_{r}' [R^{-}] p_1-k_{p} [P^{+}] p_2$.
The above ten linear equations can be solved for the ten unknowns
(six probabilities and four currents) as functions of the concentrations $[F]$, $[R^{-}]$,
and $[P^{+}]$.

The six states in the mechanochemical configuration space of the
swimmer correspond to only four independent conformations (Fig. \ref{fig:states}).
To study the coupling between the kinetics of the mechanochemical
cycle and the hydrodynamics of the ambient fluid, we need to
consider only the transitions between the distinct conformational
states. Assuming that the extensions of the arms are small compared
to their equilibrium lengths, it has been shown that the net swimming velocity
of the device is controlled by the surface area $A$ enclosed by the
conformational states that make the
$(1) \rightleftharpoons (2) \leftarrow (5) \rightleftharpoons (6) \rightarrow (1)$
loop (in Fig. \ref{fig:states}) and the net rate of sweeping this area $J$
\cite{RamArm08}. If we choose to make all the rates dimensionless by
$k_5$, and in particular define the dimensionless current $j=J/k_5$,
we can write the average swimming velocity of the device as
\begin{equation}
V=\frac{g D A k_5}{L^2} \;j\left(\frac{k_{f} [F]}{k_5},
\frac{k_{r} [R^{-}]}{k_5},\frac{k_{p} [P^{+}]}{k_5}\right),
\label{eq:V}
\end{equation}
where $g$ is a geometrical numerical prefactor of order unity
\cite{ag-pre}, $D$ is the diameter of the spheres, and $L$ is the
average total length of the swimmer.

For simplicity, let us define the dimensionless variables
$\alpha=k_{r}'/k_r$, $\beta=k_{p}'/k_p$, $\gamma=k_{4}'/k_4$,
$\delta=k_{5}'/k_5$, and $b=k_{4}/k_5$, as well as the dimensionless
concentrations $f=k_{f} [F]/k_5$, $r=k_{r} [R^{-}]/k_5$,
and $p=k_{p} [P^{+}]/k_5$. As argued above, we expect $\alpha$,
$\beta$, $\gamma$, and $\delta$ to be numerical factors (not too much)
larger than unity. Solving the above system of equations, we find
that detailed-balance holds (i.e. $j=0$), when
\begin{equation}
\gamma (\alpha \beta-\delta)+(\alpha \beta-\gamma \delta) (f+p+\alpha r)=0.
\label{eq:db}
\end{equation}
By tuning the concentrations $f$, $r$, and $p$ away from values that satisfy
the above equality, we can drive the system away from equilibrium and obtain
a nonvanishing swimming velocity ($j\neq 0$).

The concentration dependence of $j$ reveals a number of interesting features
in the device, as shown in Fig. \ref{fig:plots}. Figure \ref{fig:plots}a shows
that the dependence of $j$ and consequently the swimming velocity on the
concentration of $F$ adopts a Michaelis-Menten form.
%, with a corresponding Michaelis constant $K_M=k_5/k_{f}$.
This behavior, which is also observed in molecular motors \cite{howard}, is inherited
from the (multi-step enzymatic) reaction kinetics. Figure \ref{fig:plots}b shows
that for sufficiently low concentrations of $F$ and relatively large values of
$\gamma$, it is possible to have a situation where $j$ changes sign, which leads
to the reversal of the swimming velocity. This can be understood from the condition
of detailed-balance in Eq. (\ref{eq:db}) as $j$ is proportional to its left
hand side. Figure \ref{fig:plots}c shows that when the concentration of $F$ is
at its saturation limit (see Fig. \ref{fig:plots}a) the dependence of $j$ on the
concentrations of $R^{-}$ and $P^{+}$ is nonmonotonic, and suggests a
strategy to optimize the swimming velocity.

There are many efficient enzymatic reactions that could be used in practice
in such a setup. For example, Eq. (\ref{eq:h-reac}) can represent the
conversion of carbon dioxide ($F={\rm C O}_2$) (plus water) into bicarbonate
($Q^{-}={\rm H C O}_3^{-}$) and proton ($G^+={\rm H}^+$) catalyzed by
{\em carbonic anhydrase} ($h={\rm C A}$) \cite{stryer}. An example for
Eq. (\ref{eq:t-reac}) could be superoxide anion ($R^{-}={\rm O}_2{\cdot}^{-}$)
being scavenged by {\em superoxide dismutase} ($t={\rm S O D}$) by consuming
a proton ($P^+={\rm H}^+$) and producing hydrogen peroxide and oxygen
($S=\frac{1}{2}{\rm H}_2 {\rm O}_2+\frac{1}{2}{\rm O}_2$) \cite{stryer}.
Looking at Figs. \ref{fig:cycle} and \ref{fig:states}, we find that
in this example $k_5=k_{\rm CA}=6 \times 10^5$ s$^{-1}$ \cite{stryer}, while
$k_r=k_{\rm SOD}\sim 3 \times 10^9$ M$^{-1}$s$^{-1}$ \cite{SOD}. While
we need to know all of the rates to be able to perform a full calculation,
it is possible to make a rough estimate by looking at typical optimal values
of $j \sim 10^{-2}$ from Fig. \ref{fig:plots}, and the above estimate for $k_5$.
Using Eq. (\ref{eq:V}) with $D \sim 1$ nm, $A \sim 1$ nm$^{2}$, and $L \sim 10$ nm,
we find an estimate of $V \sim 100$ nm s$^{-1}$, which is comparable to the
efficient molecular motors that use protofilament tracks for their locomotion.
In this molecular realization, these enzymes could presumably play the role 
of the spheres themselves, and could be attached to one another using 
unstructured tails or covalently bonded polymeric tethers. It might also 
be feasible to attempt a larger scale setup with colloidal spheres covered with 
adsorbed enzymes and connected with polymeric tethers.

In practice it might be difficult to fully suppress other transitions, and
we might have to deal with more complicated configurational spaces
(e.g. involving other transition routes between these states) as well
as the back reactions. Moreover, realistic chemical reactions might have
more complicated kinetics (such as the four-stage kinetics of the reaction
of superoxide anion by SOD \cite{stryer} that we have simplified into two
stages in the above example). Such complications can be taken into account
in the calculation of the average swimming velocity using a similar
strategy \cite{RamArm09}, although they are not expected to change the
qualitative behavior of the system provided the detailed-balance is broken.

In addition to affecting the conformational transitions, thermal
fluctuations will also tend to randomize the orientation of the
swimmer, on time scales larger than the rotational diffusion time
$\tau_r \sim \eta L^3/(k_{\rm B} T)$, where $\eta$ is the viscosity
of water, $k_{\rm B}$ is the Boltzmann constant and $T$ is the
temperature. In addition to the rotational diffusion of the swimmer,
fluctuations in the density of the fuel molecule could also induce
anomalous fluctuations in the magnitude and direction of the
velocity. The combination of these effects could lead to a variety
of different behaviors, similar to those found in the context of
self-phoretic swimmers \cite{msd}. We note that a related enhanced
diffusion of active enzymes has recently been observed experimentally
\cite{enzyme-active}. The dependence of the velocity of the swimmer
on concentration of the fuel in Eq. (\ref{eq:V}) also suggests that
such artificial swimmers could be guided using chemotaxis \cite{jon,sen,lyderic}.

In conclusion, I have proposed a design for a molecular scale low
Reynolds number swimmer that can use the free energy gain and products of
a nonequilibrium chemical reaction and undergo conformational changes that
are non-reciprocal and therefore can lead to net swimming under low Reynolds
number conditions. While the present analysis has used a minimal kinetic model
that can achieve swimming at such small scales, the scheme can be readily
generalized to help analyze any realistic mechanochemical cycle
corresponding to more complex molecular devices that could be synthesized.

\acknowledgments

I would like to thank A. Ajdari, D. Lacoste, A. Najafi, E.
Rapha\"el, and H. A. Stone for fruitful discussions. This work was
supported by CNRS and EPSRC.


\begin{thebibliography}{99}
\bibitem{note0}
The studies of Refs. \cite{phoretic,jon}
rely on coupling of chemical reactions with the mechanical motion of
fluid relative to the colloid, and are in contrast with what I term as
``mechanochemical'' here, which refers to change in the morphology of
the swimmer due to chemical reactions.
\bibitem{note1}
For example, the transitions $(6) \rightarrow^{k_5} (1)$ and
$(5) \rightarrow^{k_5'} (2)$ both involve releasing a $Q^{-}$ from $h$, but
$k_5'$ corresponds to the case where the tail is in $tR^{-}$ form, which should
facilitate the release due to electrostatic repulsion as compared to when the tail
is unoccupied; thus we expect that $k_5'>k_5$.
\end{thebibliography}
\end{document}